\documentclass[11pt]{book}
\usepackage{Wiley-AuthoringTemplate}
\usepackage{chapterbib}
\usepackage[sectionbib,authoryear]{natbib}

\usepackage{graphics}	
\usepackage{amsmath}	
\usepackage{amssymb}	

\newcommand{\lapprox} {\, \lower3pt\hbox{$\sim$}\llap{\raise2pt\hbox{$<$}}\,}
\newcommand{\gapprox} {\, \lower3pt\hbox{$\sim$}\llap{\raise2pt\hbox{$>$}}\,}

\definecolor{mrkred}{RGB}{160,0,0}

\definecolor{mrk}{RGB}{0,0,160}

\begin{document}

\chapter[Helicity-conserving relaxation in unstable and merging twisted magnetic flux ropes]{Helicity-conserving relaxation in unstable and merging twisted magnetic flux ropes}

\author*[1]{Browning P.K.}
\author[1]{Gordovskyy M.}
\author[2]{Hood A.W.}

\address[1]{\orgdiv{Department of Physics and Astronomy}, 
\orgname{University of Manchester},   \street{Oxford Rd}, \city{Manchester},
\postcode{M13 9PL}, 
     \country{UK}}%

\address[2]{\orgdiv{School of Mathematics and Statistics}, 
\orgname{University of St~Andrews},  \city{St Andrews},
\postcode{KY16 9SS}, \countrypart{Scotland}, 
     \country{UK}}%

\address*{Corresponding Author: P.K. Browning} \email{philippa.browning@manchester.ac.uk}

\maketitle

\begin{abstract}{}
Twisted magnetic flux ropes  are reservoirs of free magnetic energy. In a highly-conducting plasma such as the solar corona,   energy release through multiple  magnetic reconnections can be modelled as a helicity-conserving relaxation to a minimum energy state.  One possible trigger for this  relaxation is the ideal kink instability in a twisted flux rope. We show that this provides a good description for confined solar flares, and develop from idealised cylindrical models to realistic models of coronal loops. Using 3D magnetohydrodynamic simulations combined with test-particle simulations of non-thermal electrons and ions, we predict multiple  observational signatures of such flares. We then show how interactions and mergers of flux ropes can  release free magnetic energy, using  relaxation theory to complement simulations of merging-compression formation in spherical tokamaks and heating avalanches in the solar corona.
\end{abstract}


\section{Introduction}

Magnetic flux ropes – twisted bundles of magnetic field lines – are ubiquitous structures in laboratory, space and astrophysical plasmas.
An idealised theoretical flux rope comprises a cylindrically-symmetric magnetic field with axial ($B_z$) and azimuthal ($B_{\theta}$) components (Priest, 2014), but of course, this is only an approximation to  actual structures: for example, in magnetically-confined fusion devices, twisted magnetic fields are generally bent into tori, whilst astrophysical flux ropes are 3D and often  embedded within magnetic fields of complex topology. 

Flux ropes carry electrical currents, and hence are non-potential magnetic fields and reservoirs of free magnetic energy. This feature of flux ropes is the main focus of this review, and we will show how release of this stored magnetic energy can lead to heating of the plasma and acceleration of charged particles, with application  to solar flares and solar coronal heating, as well as the formation of spherical tokamaks in the laboratory.  Flux ropes play a very  important role in eruptions in the solar corona (Liu, 2020; Patsourakos et al, 2020).  The concept of a relaxation to a minimum energy state with observed magnetic helicity provides a powerful framework for predicting energy release and for interpretation of numerical simulations and observed phenomena, with wide relevance.

Magnetic helicity is defined as  
\begin{equation}
K=\int_V \mathbf{A \cdot B} dV  
\label{helicity}
\end{equation}
where $\bf{A}$ is the vector potential corresponding to the magnetic field ${\bf B}$. As discussed in (for example) Berger (1999) and elsewhere in this volume, helicity quantifies the interlinkage of magnetic flux, or  perhaps more usefully when considering flux ropes, the overall “twistedness” of the magnetic field.  It is well known  that $K$ is  conserved  in ideal magnetohydrodynamics (MHD) (the limit of zero resistivity). In fact, the helicity of every closed magnetic flux tube is invariant - related to the fact that K is a topological property, conserved when  magnetic field lines are frozen to the plasma. Building on earlier work by Woltjer (1958), Taylor (1974)  proposed that in a "real" plasma with small but finite resistivity, the only surviving invariant from this infinite set is the total helicity, in which the integral in  equation (\ref{helicity}) is taken over the whole volume. Taylor further hypothesised that in a turbulent  plasma, subject to multiple small-scale magnetic reconnections, the plasma will relax to a state of minimum magnetic energy whilst conserving helicity. The minimum energy or relaxed state can be shown to be a linear or constant-$\alpha$ force-free field
\begin{equation}
\nabla \times \mathbf{B} = \alpha \mathbf{B}
\label{lfff}
\end{equation}
where $\alpha =\mu_0 \mathbf{j} \cdot \mathbf{B}⁄B^2$ , the ratio of parallel current to magnetic field, is spatially-uniform. This a special case of a force-free field
$\mathbf{j} \times \mathbf{B}=0$,
which can also be described by equation (\ref{lfff}) but with $\alpha$ being in general a function of position, though constant along each field line. 

Taylor’s hypothesis – as it is generally known – was successfully applied to explain the appearance of the eponymous layer of reversed toroidal magnetic field in Reverse Field Pinch devices (Taylor, 1974), and to many other laboratory plasmas (Taylor, 1984, 2000).  Helicity conservation during a relaxation event on a Reverse Field Pinch has been experimentally demonstrated by Ji et al (1995). Relaxation theory has also been successfully applied to  spheromaks (Bellan, 2000) and spherical tokamaks whilst localised relaxation models can be used to predict the fields following tokamak sawteeth crashes and Edge Localised Modes (Gimblett, 2006; Liang et al, 2010). The theory has also been widely applied to the solar corona and other astrophysical phenomena, as discussed below, and provides a nice example of the synergies between laboratory and astrophysical plasma physics. 

In order for Taylor’s hypothesis to apply, magnetic helicity must be “better” conserved than magnetic energy in the highly-conducting plasma environment of the solar corona – noting that in ideal MHD, both quantities are conserved. This will be the case if dissipation is mainly concentrated within small-scale structures (current sheets) associated with magnetic reconnection (Browning et al, 2008). More rigorous limits on the relative magnitudes of helicity and energy dissipation in the solar corona are provided by Berger (1986), and observational evidence for Taylor relaxation in the flaring  corona is given in Nandy et al (2003) and Murray et al (2013)

One of the major unsolved problems in solar physics is to explain the existence of the hot corona, whose temperature ($\approx 10^{6}$ K) is significantly higher than the photosphere  ($6000 K$) (e.g.  Klimchuk, 2006, Parnell and De Moortel, 2012; De Moortel and Browning, 2015). Coronal heating is associated with the transfer and subsequent dissipation  of magnetic energy from the solar interior into  the corona, either through MHD waves or through magnetic reconnection. The latter scenario can be considered as the combined effect of many small flare-like energy releases known as “nanoflares” (Parker, 1988). Heyvaerts and Priest (1984) first suggested that Taylor’s hypothesis could be applied to this problem. They proposed that slow motions of the photospheric footpoints of the coronal magnetic energy cause the coronal magnetic field to evolve through a sequence of force-free fields with  free magnetic energy. The stressed field then relaxes, conserving magnetic helicity but releasing free magnetic energy, to a constant-$\alpha$ state (\ref{lfff}). A series of such relaxation events may heat the corona; but in order for heating to be effective, there must be significant departures from the relaxed state. This  raises the important question of how far can the free energy build up before a relaxation event occurs: or, what triggers relaxation? One possible answer is that relaxation is triggered by the onset of an ideal instability, such as the kink instability in a twisted flux rope. This will be discussed  in the next section.

\section{Kink-instability and  relaxation in the solar corona}

A  flux rope is unstable to the ideal kink instability if the field line twist (the angle rotated by the field lines from one end of the rope to the other) exceeds a critical value: for  a periodic cylinder this is  the well-known Kruskal-Shafranov limit of 2$\pi$. For solar coronal loops  the coronal magnetic field is anchored in the dense photosphere and the resulting line-tying of the footpoints is stabilising. Hood and Priest (1979) found the critical twist thus to be somewhat larger, up to $6\pi$, dependent on the current profile. The resistive kink instability becomes unstable at a slightly lower threshold, but this grows very slowly in the highly-conducting corona, and thus the ideal kink is likely to be of far more importance as a trigger for energy release. There are many observations of the ideal kink instability in the solar corona, apparently leading to both eruptive and non-eruptive events (e.g. Srivastava et al, 2010; Yang et al, 2016). The development of the kink has also been widely studied in laboratory astrophysics experiments (e.g. Bellan, 2018). 

\subsection{Relaxation in kink-unstable flux ropes}
As discussed above, the idea of helicity-conserving relaxation to a minimum energy state can be very useful for  prediction of  the energy release in a solar flare or smaller events (nanoflares) associated with coronal heating.  In order to know how much magnetic energy is available for release, we must first know the magnetic field configuration when relaxation starts. Indeed, in order for significant free energy to be available, the field must depart significantly  from a constant-$\alpha$ state, and  relaxation can occur only intermittently. Browning and Van der Linden (2003) suggested that one likely  trigger for relaxation is the onset of the ideal kink instability in a twisted magnetic flux rope. This was motivated both by laboratory experiments  in a spheromak showing that the kink instability plays a crucial role in relaxation (Duck et al, 1997) and MHD simulations demonstrating the formation of a helical current sheet in the nonlinear phase of the ideal kink instability (Baty and Heyvaerts, 1996; Lionello et al, 1998; Gerrard et al, 2001). Vorticial motions in the photosphere slowly twist the flux rope until the instability threshold is reached; the flux rope then  rapidly develops a strong helical distortion, on Alfvenic time scales, and  a current sheet forms, allowing  magnetic reconnection and energy dissipation.  The kink instability has been shown to play a significant role in the onset of solar eruptions (e.g. T{\"o}r{\"o}k et al, 2004; Williams et al, 2005) but here we focus rather on energy release within flux ropes which do not erupt.

Browning and Van der Linden (2003) applied this concept to a family of nonlinear force-free field profiles in a cylinder, with piecewise constant $\alpha$(r). Considering finite length flux ropes with line-tied boundary conditions, the energy release was calculated at the threshold for onset of the ideal kink instability, assuming that the field relaxes to a constant-$\alpha$ field with the same helicity. This is taken to be a single heating event or nanoflare. The energy release varies over a large range, depending on the specific current profile $\alpha$(r) at instability onset. Importantly, this model predicts a minimum nanoflare size, for given loop parameters: taking a loop of length 20 Mm, radius 1 Mm and longitudinal magnetic field 30 G gives an minimal or elemental nanoflare of about $10^{18}$ J, with a possibility of  events of 10 -100 times bigger within the same loop. 

This model was extended  by Bareford et al (2010, 2011)   to determine the distribution of nanoflare energies arising from random photospheric twisting motions. Considering a family of force-free fields having piece-wise constant-$\alpha$ profile, with $\alpha=\alpha_1$ in the inner part of the loop, and $\alpha=\alpha_2$ in the outer part, it was found that the kink instability threshold forms a closed curve in $(\alpha_1,\alpha_2)$ parameter space. Slow photospheric motions cause the field profile to “wander” within this space, at some point reaching the instability threshold and then undergoing helicity-conserving relaxation – with  a release of stored energy -  to a  constant-$\alpha$ state. The whole process repeats many times, generating a distribution of nanoflare energies. For a loop of given dimensions and field strength, this produces a range of nanoflare sizes over more than one order-of-magnitude; in the case of zero-net current loops, a power law energy distribution is found (Bareford et al, 2011). A further key feature of this model is that it predicts an average shear (ratio of horizontal field to vertical field) of about 0.5, consistent with the required limit to achieve an adequate energy flux for observed coronal heating (Parker, 1988). Any successful theory of coronal heating must account for this necessary build up of free energy as well as for energy dissipation.

\begin{figure}
\centering{\includegraphics[width=1.0\textwidth]{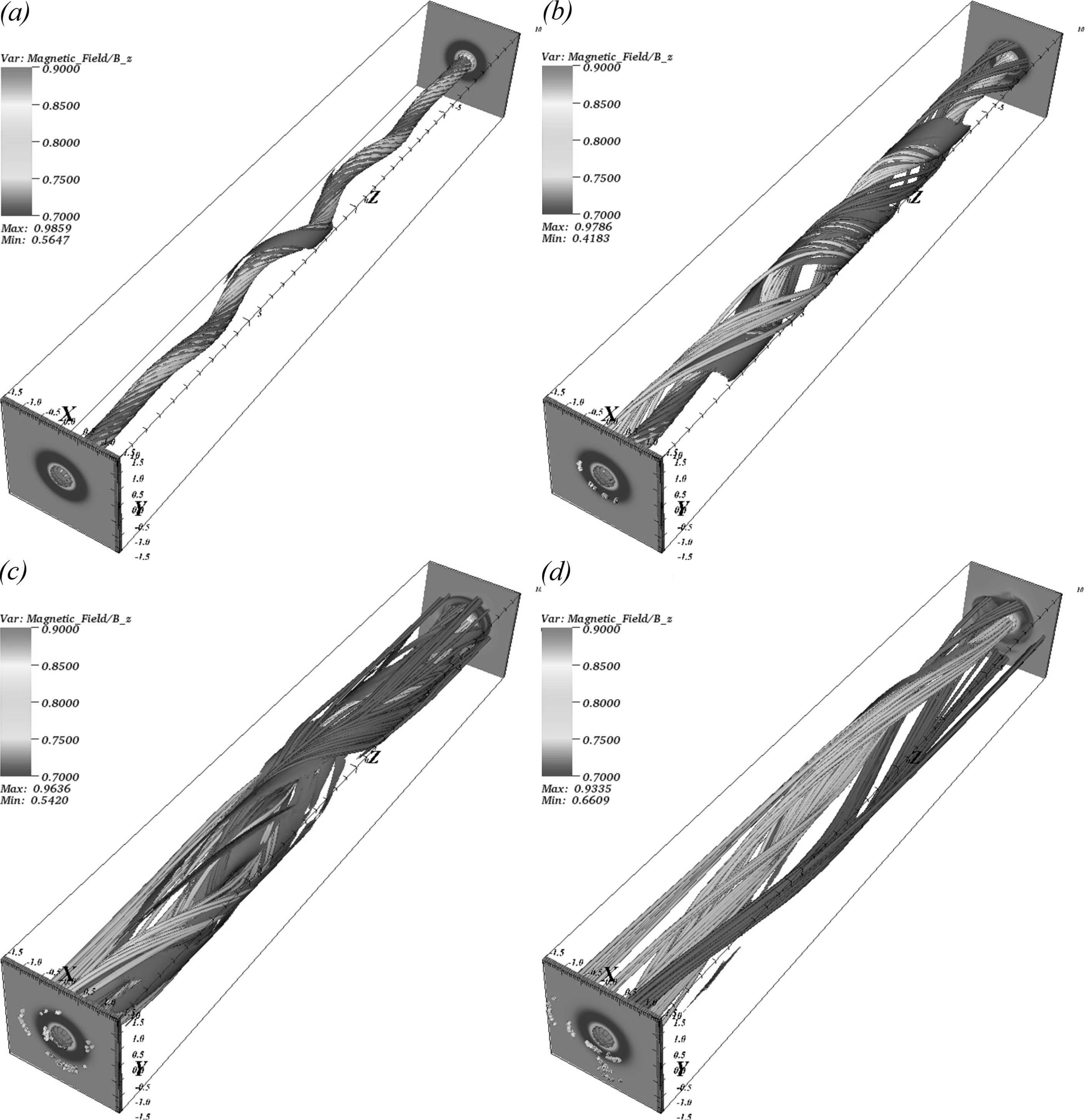}}
\caption{Magnetic field evolution in a kink-unstable cylindrical flux rope with zero net-current, from Hood et al (2009), at normalised times $t=$ 55 (top left), 85 (top right), $150$ (bottom left), $295$ (bottom right). The field lines are colour-coded by their starting locations at one end of the loop, and the dark grey  surface denotes a current  isosurface.}
\label{fig1}
\end{figure}

In order to test the hypothesis of  helicity-conserving relaxation, and also to determine the actual mechanisms by which relaxation occurs, Browning et al (2008) and Hood et al (2009) performed 3D resistive MHD simulations using the LARE3D code (Arber et al, 2000), for cylindrical loops in an initially kink-unstable equilibrium, respectively with and without net current. These simulations revealed that in the nonlinear phase of the kink, a helical current sheet formed (as seen previously), but that this subsequently fragments into multiple current sheets distributed throughout the loop volume, as shown in Figure \ref{fig1}. The splitting of the current sheet is likely to be  a 3D analogue of the well-known plasmoid instability.  Magnetic reconnection develops within these current sheets, leading to turbulent reconnection in which  outflows from each  current sheet drive further reconnections. As reconnection occurs at multiple locations, there is considerable mixing of the field lines,  flattening the $\alpha$ profile and causing the field to evolve towards the constant-$\alpha$ state. Magnetic reconnection dissipates magnetic energy, initially leading to a rise in kinetic energy associated with reconnection outflows, but eventually heating the plasma as the field settles down to a new (lower energy) equilibrium. In the zero-net current case, in which the loop is embedded within untwisted ambient field, reconnection between the loop and its surroundings also causes the loop to expand radially as it relaxes. 

These simulations show that magnetic helicity is much better conserved than energy, as expected, although the helicity dissipation is not entirely negligible due to the relatively coarse numerical resolution. The spatial distribution  of $\alpha({\mathbf r})$ in the final state is not obviously constant, reflecting a lot of remaining small-scale structure in the current; however, the magnetic field is well-approximated by a constant-$\alpha$ state, with a value of $\alpha$ which in most cases is close to that predicted through helicity conservation. Furthermore the energy release is well predicted by relaxation theory, although the actual energy released in most cases is somewhat lower than predicted by relaxation theory, presumably as a fully relaxed state is not attained by the time the simulations are ended. This confirms the validity  of Taylor’s hypothesis for calculating the energy release, justifying the use of  a semi-analytical approach which is particularly powerful for investigation of large parameter spaces. For example, Bareford et al (2010, 2011) use relaxation theory to  calculate the energy releases for $10^{5}$ different initial unstable equilibria, a task which would not be feasible with 3D numerical simulations. 

Simulations with more complex initial states consisting of braided magnetic fields have been shown to relax to  a state somewhat different from a Taylor relaxed state, consisting approximately of two very weakly twisted flux ropes with opposite signs of $\alpha$. This has been interpreted as being due to the presence of additional topological constraints not included in Taylor’s hypothesis (Yeates, 2015). On the other hand, it should be noted that there is no expectation within the framework of relaxation theory that a minimum energy state should {\it always} be attained: this should only occur if there is a sufficiently high level of turbulent reconnection to allow most field lines to reconnect, otherwise relaxation may not occur at all, or it may cease at some meta-stable state with energy higher than the minimum. Indeed, the case of two flux ropes with opposite twist is a good example of a state with free magnetic energy, but in which the reconnection required to   access to the minimum energy state (a potential field in this case) is unlikely to occur, since the azimuthal fields at the interface between the flux ropes are in the same direction. 

It should also be noted that relaxation theory has to be modified when considering astrophysical plasmas, since (unlike laboratory plasmas) there are no conducting walls to define the extent of relaxation. Therefore, we must have {\it partial relaxation}, in which the field relaxes over a volume of limited extent, whilst the surrounding magnetic field remains undisturbed (Bareford et al, 2013). The extent of the partial relaxation region depends on the turbulent reconnection driving the relaxation. In a kink-unstable loop, this is  controlled by the nonlinear amplitude of the unstable mode; typically the relaxed loop has radius around 1.5 times its initial value. We must rely on numerical simulations to determine this parameter- however, the energy release is not highly sensitive to this (Bareford et al, 2013). 

\subsection{Observable signatures of energy release in flux ropes}
We have considered so far a rather idealised cylindrical geometry, whereas coronal loops have both global curvature and field expansion. Also, it has been assumed that the initial state is unstable to the kink instability, whereas in reality, a loop must be slowly twisted across the kink instability threshold. We now describe more realistic models, which furthermore can properly determine the plasma heating by incorporating radiation and thermal conduction in the energy equation, which we now describe.

One of the major questions in solar flare physics is to explain the origin of the large numbers of non-thermal electrons and ions, which carry a significant fraction of the released energy. Whilst the electric fields associated with magnetic reconnection can accelerate particles, it is difficult to accelerate sufficiently large numbers of particles within a narrow current sheet, and it has thus been suggested that distributed acceleration throughout a large volume of the flaring region is more feasible (e.g. Cargill et al, 2012).  The scenario of a kink-unstable flux rope provides a means to do this, due to the fragmented nature of the current sheets. 

Using test particles combined with 3D MHD simulations, Gordovskyy and Browning  (2011, 2012) showed that 5-10 \% of protons and electrons could be accelerated out of the thermal distribution, with energies ranging from about 1 keV up to 1 MeV, and hard power-law like spectra with indices 0.8 – 1.5. The particles gain energy in a stepwise manner as they repeatedly encounter current sheets, and the loop quickly fills with energetic particles. Acceleration is predominantly through the parallel electric field in the current sheets, and hence the pitch angle distribution of the energetic particles is mainly parallel. 

\begin{figure}
\centering{\includegraphics[width=1.0\textwidth]{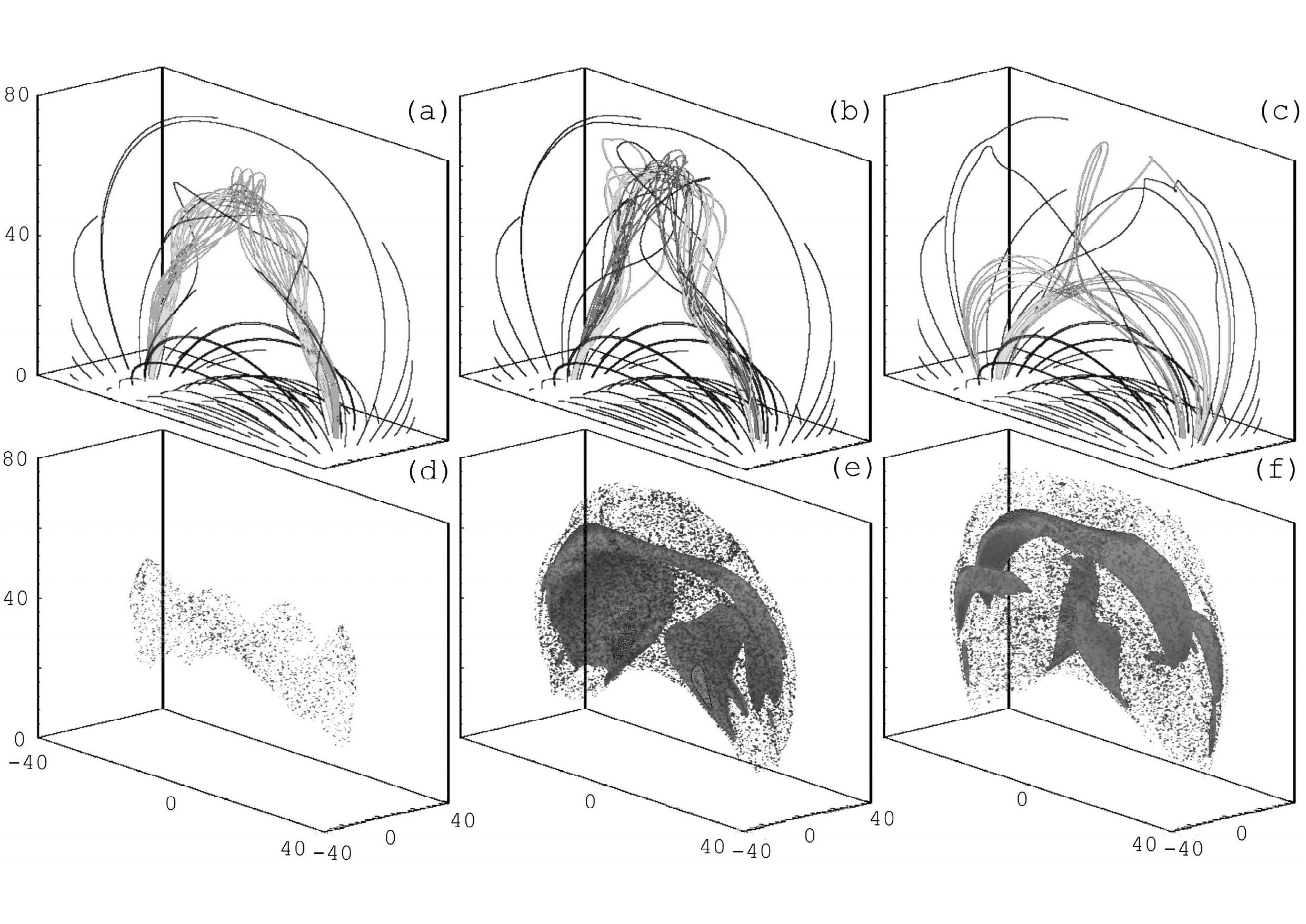}}
\caption{Evolution of a reconnecting twisted loop in gravitationally stratified corona (based on Gordovskyy et al. 2014, 2017). Panels (a-c) show magnetic field lines, different shades  used to demonstrate the change of connectivity. Panels (d-f) show distribution of the hot plasma, with dark grey patchy surface corresponding to 2MK, mid grey and light grey  surfaces corresponding to 3 and 4MK, respectively. Panels (a) and (d), (b) and (e), and (c) and (f) correspond to times of about 15s, 43s and 72s after the kink, respectively. Axis units are in megameters.}
\label{loop-mhd}
\end{figure}

\begin{figure}
\centering{\includegraphics[width=0.5\textwidth]{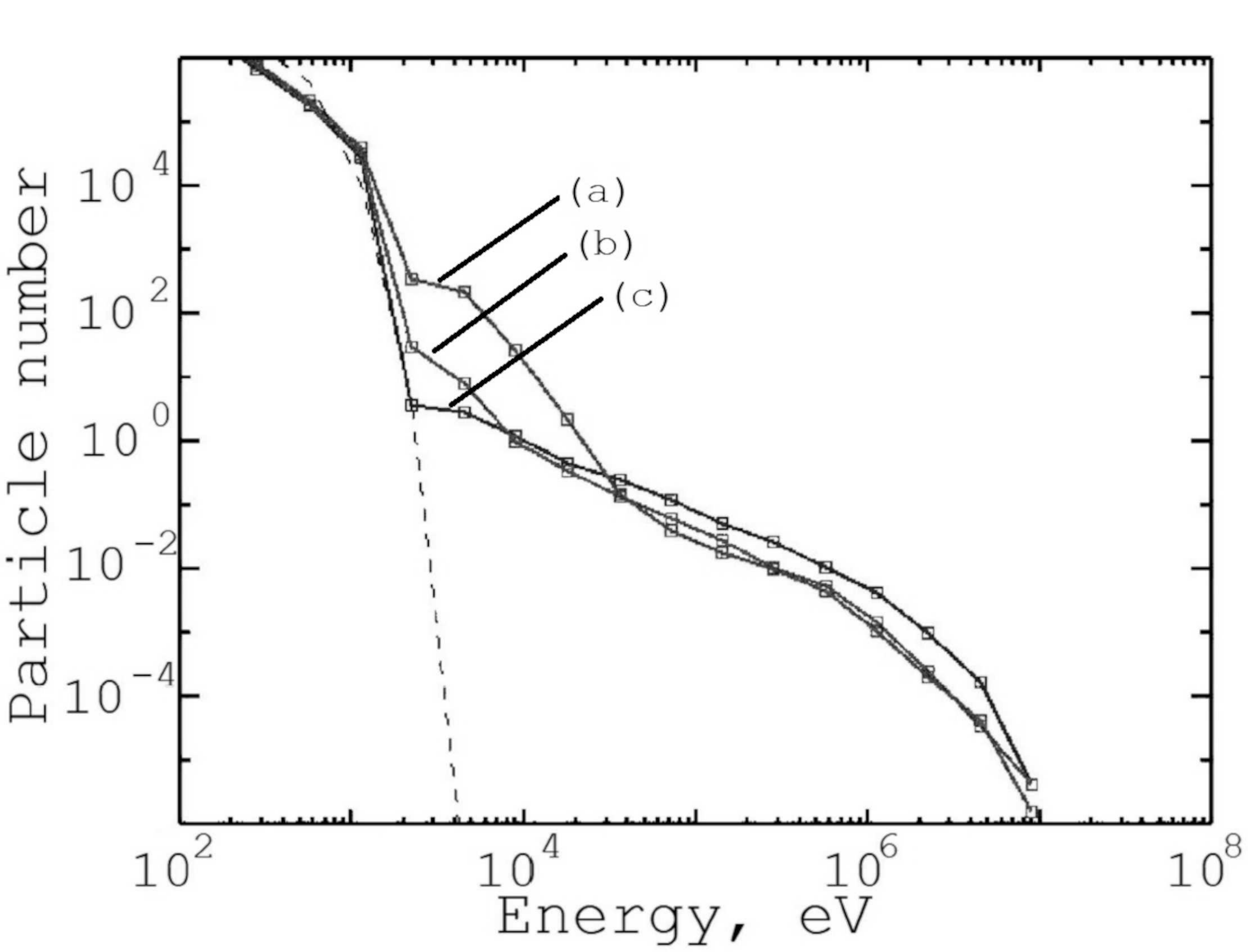}}
\caption{Electron energy spectra in the model of a reconnecting twisted loop (based on Gordovskyy et al. 2014). Lines (a), (b), and (c) correspond to approximately 15s, 30s and 120s after onset of instability. The dashed line denotes the initial Maxwellian distribution}
\label{loop-particles}
\end{figure}

One important feature of “real” coronal loops compared with cylinders is magnetic convergence – the field  lines converge towards the photospheric footpoint sources, and the field strength is thus significantly weaker at the looptop than the footpoints. The effects of this on a kink-unstable flux rope have been  addressed by Gordovskyy et al (2013, 2014), as well as the effects on particle energisation  of Coulomb collisions with the  background plasma (Figures~\ref{loop-mhd}-\ref{loop-particles}). There is no useful analytical form for the equilibrium, so the twisted loop is formed numerically by slow rotational motions of the footpoints, until it becomes kink unstable. At this point, the loop behaves similarly to the earlier cylindrical models: kinking, forming fragmented current sheets and   relaxing to a state of lower magnetic energy. The magnetic convergence somewhat changes the spatial distribution of accelerated electrons, as electrons are more strongly accelerated in the footpoint regions due to the stronger currents associated with field line convergence.  Collisions with the background plasma are much more frequent in the dense plasma layers near the footpoints (the chromosphere) – and the resultant pitch-angle scattering there enhances the trapping of energetic electrons in the loop, since they are mostly scattered out of the loss cone.

The combination of test particles with 3D MHD simulations  - the MHDTP approach - provides a powerful basis for forward modelling of both thermal and non-thermal emission from solar flares (Gordovskyy et al, 2019).  In order to model properly the distribution of energetic particles, the population of test particles must be tracked through  the time-evolving electromagnetic fields, interpolating the fields in space and time to the particles’ locations (Gordovskyy et al, 2010). 
3D MHD simulations of  a confined solar flare in a twisted coronal loop are performed by slowly twisting  the footpoints of a loop within a dipolar magnetic region, embedded in a gravitationally-stratified atmosphere  until an instability sets in. The subsequent evolution is qualitatively similar to earlier studies of cylindrical loops (Bareford et al, 2016), but the curved loops usually also undergo an upwards  "failed eruption" and large-scale reconnection before relaxing to a lower-energy weakly-twisted loop  as shown  in  Figure~\ref{loop-mhd}  (Bareford et al, 2016, Pinto et al, 2016). The release of stored magnetic energy provides significant plasma heating. However, since the soft-Xray/EUV emission from this heated plasma  is mainly visible after the reconnection has significantly reduced the field line twist, the initial strong twist of the field lines is not visible in the thermal  emission (Pinto et al 2015, Pinto et al 2016). Thus, other observational features must be used to confirm the presence of twisted magnetic fields. 

The fragmented current sheets in the kink-unstable flux rope allow effective acceleration of particles throughout the loop volume, leading to  strong non-thermal tails in the particle energy distributions (Figure \ref{loop-particles}). The MHDTP approach in the twisted loop scenario has allowed forward modelling and analysis of a range of observational signatures in both thermal and non-thermal emission, as follows. Hard X-ray spatial locations show  footpoint sources growing  in time,  and sometimes looptop or extended loop  emission (Gordovskyy et al, 2014). Predicted soft X-ray and hard X-ray light curves (Pinto et al, 2016) are broadly consistent with observations, showing spikey and transient non-thermal emission, with smoother and more protracted thermal emission. The average line-of-sight velocities and velocity dispersion, leading to non-thermal spectral  line broadening, has been investigated in a relaxing twisted loop (Gordovskyy et al, 2016), showing positive correlations between velocity dispersion and temperature consistent with observations. 

Perhaps the  most specific signature of energy release within twisted magnetic fields comes from polarisation of microwaves created by gyrosynchrotron emission from mildly relativistic electrons. It has been shown (Gordovskky et al, 2017) that the characteristic patterns in microwave polarisation characteristic of  twisted magnetic magnetic fields  are most likely to  be observable in higher microwave frequencies ($> 30 $GHz), and  for loops seen on the limb; also,  the effect is strongest  in the initial phases of reconnection and energy release, when the number on non-thermal electrons is greatest. 

Studies in more idealised cylindrical geometry have also  predicted observational signatures in soft X-rays (Pinto, 2015), Hinode EIS emission lines (Snow et al, 2017), notably loop expansion and intensity enhancements at the loop edge, as well as multiple signatures potentially observable with DKIST (Snow et al, 2018).

We may summarise the predicted observable signatures of energy release in reconnecting twisted loops, and corresponding  observational evidence as follows:
 \begin{itemize}
    \item Expanding footpoint Hard X-ray Source. Some observational evidence  (e.g. Kontar et al, 2011). 
    \item Weakly-twisted filamentary extreme UV and  soft X-ray emission, with the apparent twist angle being significantly lower than the twist angle expected  for the kink instability  3.5$\pi$.  Widely observed e.g. Srivastava et al., 2009, but not necessarily characteristic of this scenario. 
    \item Spiky and transient Hard X-ray light curves, with more prolonged soft X-ray emission. A general feature of observed flares, but not specifically characteristic of this scenario 
    \item Correlation of velocity dispersion (non-thermal line-broadening) with temperature, and other features in flow patterns. Agrees with observations very well (e.g. Doschek et al, 2008), but this behaviour may be typical of plasma heating with turbulent reconnection outflows, not necessarily in twisted fields.
    \item Microwave polarisation patterns. Some observations (e.g. Sharykin et al, 2018), with potential for future observations with high-resolution facilities.
    \item Actual or failed eruption. (By failed eruption we mean initial expansion of the twisted loop after onset of energy release, followed by the contraction of the loop. Both types of coronal loop eruption are widely observed e.g. Leaman et al., 2003, Alexander et al., 2006, Srivastava et al. 2009, Kuridze et al. 2013).
\end{itemize}
Along with the footpoint expansion, cross-loop microwave polarisation gradient can be considered key observational signatures  indicating energy release and particle acceleration in a twisted coronal loop. 
Note that no single observation could provide  conclusive evidence of energy release in twisted loops rather a combination of multi-wavelength observations is needed. Many of these signatures are characteristic of energy release in distributed reconnection sites, which can arise in different flaring scenarios and not just in twisted fields; indeed, the methodology developed here could very usefully be applied to configurations beyond the single twisted loop scenario. However, energy release within a twisted loop does provide an attractive scenario for confined flares in which a loop transiently brightens but does not erupt, as well as so-called "failed" eruptions, and the modelling described  here may provide a guide for interpretation of observations, as well as suggesting observations for new instruments.

\section{Merging flux ropes}

\subsection{Flux rope merging in spherical tokamaks}
So far, we have discussed  only individual twisted loops. However, fields with multiple twisted flux ropes may contain free magnetic energy which could be released as the flux ropes merge through magnetic reconnection (Browning et al, 1986). As twisted flux ropes merge into a smaller number of ropes, the typical scale-length of the field increases: this may be  interpreted as an inverse-cascade of energy from small scales to large scales, or as self-organisation process.  Reconnection and energy release in merging flux ropes is likely to be important in the solar corona, and has been  modelled (e.g. Linton et al, 2001; Kliem  et al, 2014)  as well as being observed 
in many eruptive events, see Liu (2020)  and references therein. Furthermore, merger of flux ropes may play a role in the pre-eruption formation of a large-scale flux rope (Patsourakis et al, 2013, 2020; Kliem et al 2021). The merger of magnetic flux ropes has also been studied in a number of dedicated laboratory reconnection experiments, but here we consider reconnection in a spherical tokamak which, due to the high plasma temperature, is in a parameter  regime more closely akin to the solar corona than most dedicated  experiments. The plasma $\beta$ (around $10^{-4}$ -  $10^{-2}$) is comparable to the solar corona, and the Lundquist number $S$ is also large, up to $10^{7}$ (Browning et al, 2014). 

Merger of two  flux ropes has been classified as “co-helicity”  or “counter-helicity”(Yamada et al, 1990), depending on whether, respectively, the flux rope helicities are of the same or opposite signs. We focus here mainly on the former, co-helicity, case, which is found in the spherical tokamak and seems more likely in the solar corona. In this case, the system initially has a non-zero helicity which provides a constraint on the relaxation, the final state being a single twisted flux tube. In the counter-helicity case, the initial total helicity is zero, so the field could  relax to a potential state, releasing more free magnetic energy.

The spherical tokamak is a fusion device consisting of a toroidal magnetic configuration with very low aspect-ratio: essentially, it is a conventional tokamak with no central hole, instead having a central column carrying the toroidal field current and also (often) a solenoid generating poloidal field. There is considerable interest in developing a start-up scheme not involving the central solenoid, and merging-compression, in which two current-carrying plasma tori merge, is an attractive scenario which has been explored in the MAST spherical tokamak and other devices (Gryaznevich and Sykes, 2017). Ramping down the current in poloidal field coils inductively forms two current-carrying flux ropes, which are attracted by the mutual currents, thus detaching from the field coils and moving together. The two flux ropes merge into a single twisted flux rope, with significant  heating, thus forming a hot spherical tokamak plasma  (see, for example, Figure 1 of Stanier et al (2014) for a sketch of the field and plasma geometries. 

Stanier et al (2013) performed two-fluid and resistive MHD simulations of the merging-compression process in MAST. The initial state comprises a superposition of two current-carrying flux ropes which are seperately in force-free equilibrium but experience an attractive Lorentz force due to their parallel currents. The flux ropes move together, forming a current sheet and reconnecting, with some oscillations, eventually forming a single flux rope with a spherical-tokamak magnetic  field configuration (in low aspect-ratio simulations). The magnetic reconnection is associated with significant plasma heating, and the temperature profiles in the two-fluid simulations are in good agreement with measured ion and electron temperatures (Browning et al, 2016). The flux rope merging is also associated with particle acceleration, with observations of high-energy ions, and more indirect evidence for electrons, as predicted by modelling of the reconnecting fields in MAST (McClements, 2019).

\begin{figure}
\centering{\includegraphics[width=1.0\textwidth]{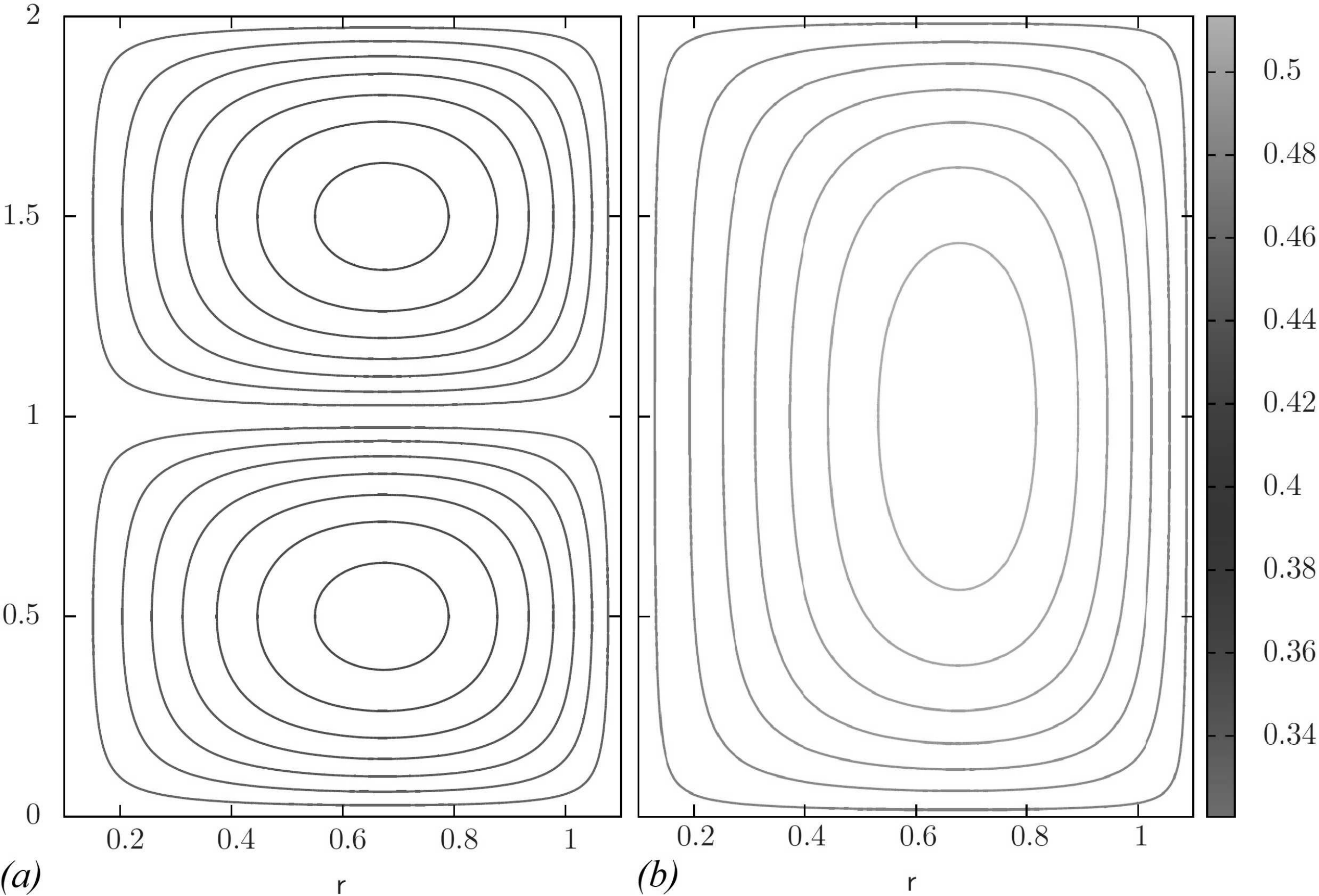}}
\caption{Poloidal field lines for the  relaxation model of merging flux ropes in a tight-aspect ratio rectangular container, from Browning et al (2016). The initial flux ropes (left ) each have $\alpha H = 1.25$, while the final relaxed state (right) has  $\alpha H = 0.849$, where $2H$ is the container height.  }
\label{merging}
\end{figure}

This process can be usefully modelled as a Taylor relaxation to a minimum energy state (\ref{lfff}) (where in the fusion literature, $\alpha$ is denoted by $\mu$). A semi-analytical model has been developed for a plasma contained within a conducting chamber of rectangular cross-section,  in both infinite-aspect ratio geometry (Browning et al, 2014) and in a tight-aspect ratio cylindrical configuration (Browning et al, 2016). The initial field is modelled as two adjacent  flux ropes of rectangular cross-section; each flux rope is a constant-$\alpha$ field but a current sheet (discontinuity in poloidal field) between the flux ropes means that the system has free energy. This is assumed to relax to a final state which is a constant-$\alpha$ field throughout the region, where the value of $\alpha$ and the peak value of the field are calculated by applying the constraints of helicity and toroidal flux conservation. A typical case is shown in Figure \ref{merging}. The final state naturally has lower magnetic energy than the initial state, and the loss in magnetic energy will be converted into plasma thermal energy. Taking parameters representative of MAST, with plasma current of 300 kA in each initial flux rope, the average temperature rise is estimated to be 150 eV, consistent with observations. The free magnetic energy - and hence the final temperature are shown to have an approximately quadratic scaling with  plasma current. 

The relaxation model is potentially useful for making quick predictions of the field configurations and heating across a wide parameter space. It can also be easily adapted to model merging of coronal loops, where different initial configurations, such as multiple loops, or loops with opposite twist, could also  arise (Browning et al, 2016). For two flux ropes, the energy release is found to be substantially greater if the flux ropes initially have opposite currents (twisted in reverse senses) than if the currents are in the same direction. This is because the former configuration has zero helicity, and hence the relaxed state has lower energy. However, the oppositely-twisted flux ropes are much less likely to relax and release the free energy, since neither the toroidal or poloidal magnetic field components reverse at the interface between the flux ropes, and there is no obvious way - apart perhaps from a strong external perturbation - in which magnetic reconnection could be initiated. 
If an initial larger array of  flux ropes is considered, the energy release scales almost linearly with the number of flux ropes, since the free magnetic energy is associated with the current sheets between flux ropes. This relaxation from many  flux ropes to a single rope, releasing magnetic energy, exhibits the natural tendency for an inverse cascade from small to large length-scales. 

\subsection{Heating avalanches in the solar corona}

It has long been considered that flux rope interactions and  mergers in the solar corona may release free magnetic energy leading to a large-scale flare or smaller events associated with solar coronal heating.   Building on the modelling of individual kink-unstable flux ropes described above, Tam et al (2015) investigated the dynamics of  a pair of adjacent flux ropes, representing separately-twisted threads within a coronal loop, using 3D MHD simulations. It was discovered that if an unstable thread is located close to a stable thread, the instability in the former could trigger energy release in both threads. The threads interact through magnetic reconnection, and the final state is a single weakly-twisted flux rope.  Thus, free magnetic energy from a twisted flux rope can be released even if it is kink-stable. An alternative scenario for interacting flux ropes is considered by Ripperda et al  (2017a, 2017b), with oppositely-directed (and hence repelling) current  channels in which the tilt instability causes reconnection in its nonlinear phase. Similarly to Gordovskyy et al (2010) and subsequent work described above, the relativistic guiding-centre equations are used to model the evolution of non-thermal ions and electrons. Strong particle acceleration is shown to take place, forming particle energy distributions with significant peaks at high energies ("bump on tail" distributions).

\begin{figure}
\centering{\includegraphics[width=1.0\textwidth]{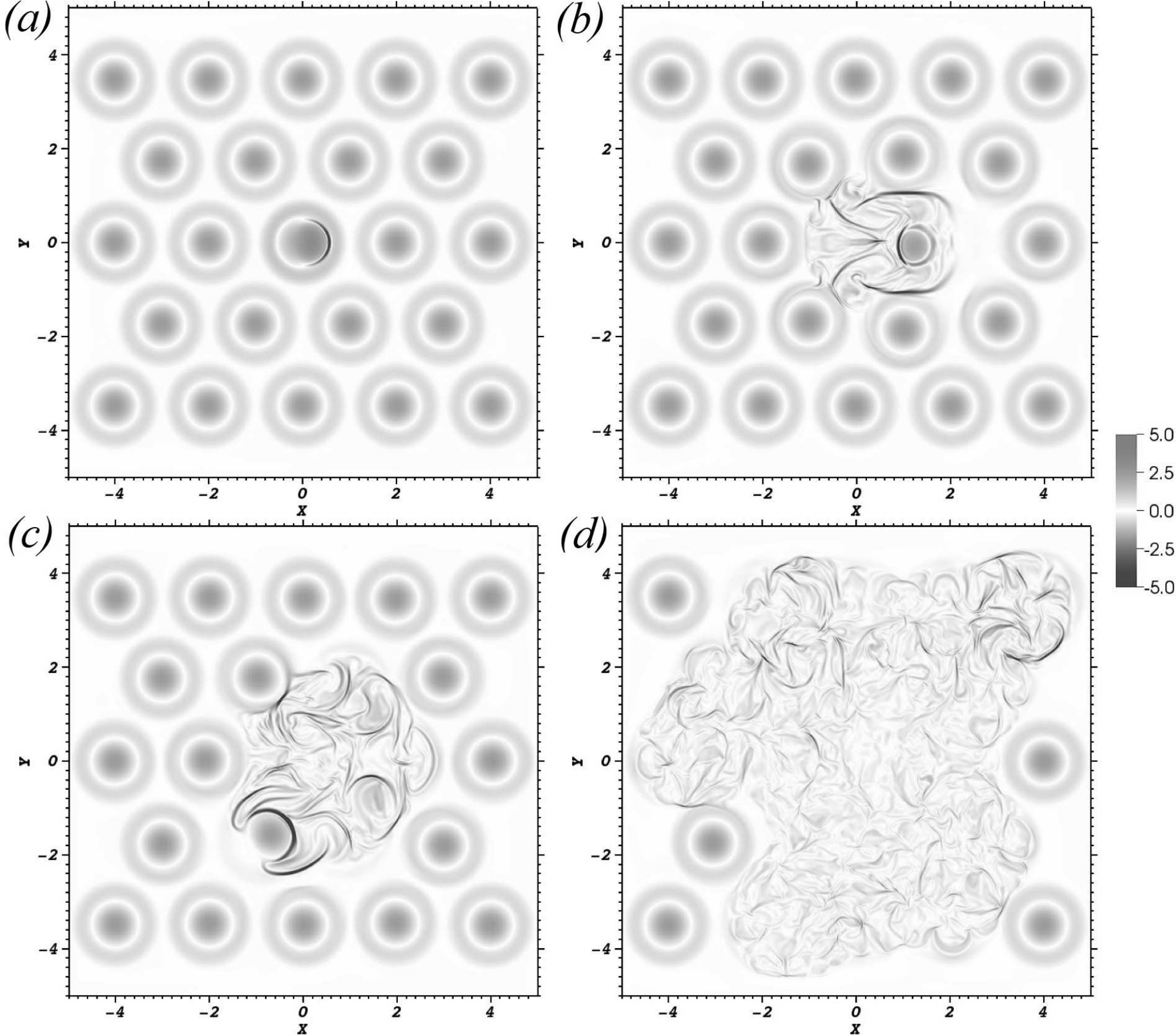}}
\caption{Contours of axial current in the midplane of an array of 23 parallel twisted threads, from Hood et al (2016), with one  initially  kink-unstable  central thread and all other threads initially stable,   at normalised times $t=$ 75 (top left), 165 (top right), $290$ (bottom left), $800$ (bottom right). }
\label{avalanche}
\end{figure}

These studies lead to the question of whether a single unstable twisted thread  could trigger energy release from multiple stable threads: Hood et al (2016) demonstrated that this could indeed be the case. They used 3D MHD simulations to investigate the evolution of  an array  of 22 stable threads, with only one unstable thread, showing that an "avalanche" of heating ensued, with the initially unstable flux rope interacting and merging successively with most of its stable neighbours (see Figure \ref{avalanche}). This was the first demonstration using a MHD model of an avalanche process, often postulated in cellular automaton models of coronal heating. This has major implications for solar coronal heating, since a localised instability can trigger heating in a much larger volume.

This scenario of merging flux ropes can be usefully  described as a helicity-conserving relaxation process (Hussain et al 2017). Two (or more) twisted flux ropes may relax to a minimum energy state of a single constant-$\alpha$  twisted flux rope, with the same total helicity as the  initial configuration but with lower energy. This semi-analytical model was benchmarked against the  MHD simulations of Tam et al (2015), giving good agreement in the values of released energy. The energy release  does depend on the radius of the relaxed flux rope, as relaxation will only extend over a limited volume with the surrounding potential magnetic field remaining undisturbed (Bareford et al, 2013); a simple conservation of volume approach gives good agreement with numerical results. The relaxation model can be applied to larger collections of flux ropes, including the specific system modelled by Hood et al (2016); the relaxation model predicts that the magnetic energy decreases in approximately steps as additional flux ropes are "consumed" in the avalanche. While 3D numerical simulations are computationally expensive, the relaxation approach allows exploration of wide parameter spaces, and opens up the possibility of developing a cellular-automaton-type model with rules for interactions and energy release developed from relaxation theory .  

However, relaxation theory itself cannot predict whether two twisted threads will or will not actually interact and relax: the conditions under which this will occur must be determined through MHD simulations. Preliminary work by Hussain (2018) suggests that relaxation is unlikely to take place between oppositely-twisted threads -  or if it does do so, will occur more slowly, and requires stronger triggering. This is because, as discussed above, there is no reversal of the poloidal field component at the interfaces between threads in this case; even though the relaxed state is closer to a potential field and thus the free energy is greater than for like-twisted threads. Furthermore, relaxation takes place if threads are sufficiently close, with a separation of less than about 0.25 - 0.5 radii required for interaction. 

The  simulations of Tam et al (2015) and Hood et al (2016) consider an initial set up with one thread in an kink-unstable state. This was extended by Reid et al (2018) to show that similar behaviour can arise  in a field which - more realistically - starts in a stable untwisted state and is continuously driven by footpoint motions until it goes unstable. As the driving is maintained, a series of bursts of heating ensue. Indeed, we expect the corona to be generally in a state of "driven relaxation", in which energy input from  footpoint motions statistically  balances energy dissipation through relaxation. Driven relaxation associated with external helicity injection is also integral to spheromaks (Bellan, 2000) and to spherical tokamaks such as NSTX in which current drive is through coaxial helicity injection. 

\section{Discussion and conclusions}
Twisted flux ropes store free magnetic energy, which may be released either in individual flux ropes in the nonlinear phase of the ideal kink instability, or through interaction of multiple twisted flux ropes. In the former case, the field  relaxes to a state of minimum magnetic energy through internal magnetic reconnections as well as  reconnection with ambient untwisted field lines. In the latter case, multiple flux ropes  reduce their energy through reconnecting into a single flux rope.  The theory of helicity-conserving relaxation to a minimum energy state - a constant-$\alpha$ force-free field (Taylor, 1974) - provides a powerful tool for modelling such processes and a framework for interpretation of observations and simulations. A key strength of this approach is that the final equilibrium state and the energy released during reconnection can often be calculated analytically, allowing exploration of very large parameter spaces which are unfeasible with numerical simulations. 

Numerical MHD simulations of unstable or merging flux ropes provide considerable insight into the process of relaxation and the underlying magnetic reconnection. Furthermore, these allow predictions of observable signatures of energy release, including both thermal and non-thermal emission (if a test-particle model for non-thermal particles is integrated with the MHD model). Energy release in twisted coronal loops provides a viable scenario for confined solar flares, with the release of stored magnetic energy leading to both plasma heating and particle acceleration. {\bf  It is required to consider the characteristics of a range of observables in both thermal and non-thermal emission in order to find clear evidence for energy release in  twisted magnetic fields in flares}. 

Comparison of 3D MHD simulations with the predictions of relaxation theory show that helicity-conserving relaxation is useful model, especially for calculation of magnetic energy changes. The energy of the final state of the magnetic field may be well-approximated by the energy of the constant-$\alpha$ state, even if  $\alpha$ is not obviously  close to being constant. The reason for this is twofold. Firstly, since energy is an integral of the magnetic field, which is itself an integral of the current, it is insensitive to fluctuations in the current. Secondly, because the relaxed state is an energy minimum, the energy deviation from this depends quadratically on  deviations in the current profile. Numerical simulations also provide insight into the mechanisms by which relaxation occurs, whereby multiple reconnections create many interchanges of magnetic field lines and hence mix the $\alpha$ profile. 

However, relaxation theory must be adapted when applied to the solar corona - and other astrophysical contexts. Relaxation can only   be partial, and the extent of the relaxed region has to be postulated using knowledge   gained from numerical simulations or some other means. For example, a zero-net current flux rope will undergo some reconnection with the ambient field and hence the relaxed flux rope  is somewhat greater  in size than the initial flux rope - but still the ambient field at larger distances does not participate in the relaxation, so relaxation is highly localised. Furthermore, the coronal field is constantly  subject to disturbances by photospheric footpoint motions, and thus there is an interplay between relaxation and driving, and a fully relaxed state can never be attained. Similar considerations apply in some laboratory plasmas. 

We have shown that helicity-conserving relaxation can be applied to both solar coronal loops and to spherical tokamaks, and that concepts and methods can be fruitfully shared between these fields. Many other  applications have been considered, some also emphasising the role of the kink instability: for example, Bromberg et al (2019) consider the kink instability in relativistic astrophysical jets, and find that the relaxed state can be described as a Taylor state.

\section*{Acknowledgments}

PKB and MG acknowledge financial support from the UK Science and  Technology Facilities Council (grant ST/T00035X/1).
AWH also acknowledges STFC (grant ST/S000402/1).

\section*{References}

Alexander, D., Liu, R. and Gilbert, H.R. (2006). What Is the Role of the Kink Instability in Solar Coronal Eruptions? Astrophys.J.Lett., 569, L255-L258, doi:10.1086/379530

Arber,T.D, Longbottom, A.W., Gerrard, C.L. and  Milne,A.M. (2001) A staggered grid, Lagrangian-Eulerian  Remap code for 3D –MHD simulations, J. Comp. Phys. 171, 151, doi: 10.1006/jcph.2001.6780

Bareford, M.R., Browning P.K and Van der Linden,R.A.M. (2010) A nanoflare distribution generated by repeated relaxations triggered by kink instability. Astron. Astrophys. 521, A70. Doi:10.1051/0004-6361/201014067

Bareford, M.R., Browning P.K and Van der Linden,R.A.M  (2011) The Flare-Energy Distributions Generated by Kink-Unstable Ensembles of Zero-Net-Current Coronal Loops. Sol. Phys. 273, 93-115, doi:10.1007/s11207-011-9832-4

Bareford,M.R., Hood,A.W. and Browning,P.K. (2013) Coronal heating by the partial relaxation of twisted loops. Astron. Astrophys. 550, A40. doi: 10.1051/0004-6361/201219725

Bareford,M.R., Gordovskyy, M. , Browning, P.K. and Hood, A.W. (2016) Energy Release in Driven Twisted Coronal Loops, Sol. Phys. 291, 187-209, doi:10.1007/s11207-015-0824-7

Baty, H. and Heyvaerts, J. (1996)  Electric current concentration and kink instability in line-tied coronal loops. Astron. Astrophys. 308, 935-950.

Bellan,P.M. (2000) Spheromaks: a practical application of magnetohydrodynamic dynamos and plasma self-organization, Imperial College Press.

Bellan, P.M. (2018) Experiments relevant to astrophysical jets. J.  Plas. Phys. 84, 755840501, doi:10.1017/S002237781800079X

Berger, M.A. (1984) Rigorous new limits on magnetic helicity dissipation in the solar corona. Geophysical and Astrophysical Fluid Dynamics, 30, 79-104. doi.org/10.1080/03091928408210078

Berger,M.A. (1999) An Introduction to magnetic helicity. Plas. Phys. Cont. Fus. 41, 167B. doi:10.1088/0741-3335/41/12B/312 

Bromberg,O., Singh,C.B, Davelaar,J. and Philippov,A.A. (2019) Kink Instability: Evolution and Energy Dissipation in Relativistic Force-free Nonrotating Jets. Astrophys. J. 884, 39. doi:10.3847/1538-4357/ab3fa5

Browning, P.K., Sakurai,T. and Priest, E.R. (1986) Coronal heating in closely-packed flux tubes: a Taylor-Heyvaerts relaxation theory. Astron. Astrophys. 158, 217-227.

Browning, P. K. and Van der Linden, R.A.M. (2003) Solar coronal heating by relaxation events. Astron. Astrophys. 400, 355-367, doi:10.1051/0004-6361:20021887.

Browning, P.K., Gerrard, C., Hood, A.W., Kevis, R. and van der Linden, R.A.M. (2008) “Heating the corona by nanoflares: simulations of energy release triggered by a kink instability", Astron. Astrophys., 485, 837-848,
doi 10.1051/0004-6361:20079192

Browning, P.K., Stanier, A., Ashworth, G., McClements, K.G. and Lukin, V.S.  (2014) Self-organization during spherical torus formation by flux rope merging in the Mega Ampere Spherical Tokamak.  Plas. Phys. Cont. Fus, 56, 064009, doi:10.1088/0741-3335/56/6/064009

Browning, P.K., Cardnell, S., Evans, M., Arese Lucini, F., Lukin, V.S., McClements, K.G. et al. (2016) Two-fluid and magnetohydrodynamic modelling of magnetic reconnection in the MAST spherical tokamak and the solar corona, Plas. Phys.  Cont. Fus.  58, 014041, doi:10.1088/0741-3335/58/1/014041

Cargill,P.J., Vlahos,L., Baumann,G, Drake, J.F. and Nordlund,A (2012) Current Fragmentation and Particle Acceleration in Solar Flares, Space Sci. Rev. 173, 223. doi:10.1007/s11214-012-9888-y

Doschek,G.A., Warren,H.P., Mariska,J.T., Muglach,K., Culhane,J.L., Hara,H. and Watanabe,T. (2008) 
Flows and Nonthermal Velocities in Solar Active Regions Observed with the EUV Imaging Spectrometer on Hinode: A Tracer of Active Region Sources of Heliospheric Magnetic Fields. Astrophys. J. 686, 1362-1371, doi:10.1086/591724.

Duck, R.C, Browning, P.K., Cunningham, G., Gee, S., al-Kharkhy, A., Martin, R.  and Rusbridge, M.G. (1997)  Structure of the n = 1 mode responsible for relaxation and current drive during sustainment of the SPHEX spheromak. Plas.Phys.  Cont. Fusion 39, 715-736. 10.1088/0741-3335/39/5/004,

Gerrard, C.L, Arber, T.D, Hood, A.W. and Van der Linden, R.A.M.  (2001). Numerical simulations of kink instability in line-tied coronal loops. Astron. Astrophys. 373, 1089-1098. 10.1051/0004-6361:20010678

Gimblett, C.G., Hastie, R.J. and Helander, P. (2006)  Model for Current-Driven Edge-Localized Modes, Phys. Rev. Lett. 96, 035006,  doi:10.1103/PhysRevLett.96.035006.

Gordovskyy,M., Browning, P.K. and Vekstein, G.E. (2010) Particle Acceleration in Fragmenting Periodic Reconnecting Current Sheets in Solar Flares, Astrophys. J. 720, 1603-1611, doi:10.1088/0004-637X/720/2/1603.

Gordovskyy,M. and Browning, P.K. (2011) Particle Acceleration by Magnetic Reconnection in a Twisted Coronal Loop, Astrophys. J. 729, 101,  doi:10.1088/0004-637X/729/2/101

Gordovskyy,M. and Browning, P.K. (2012) Magnetic Relaxation and Particle Acceleration in a Flaring Twisted Coronal Loop, Sol. Phys., 277, 299-316, doi:10.1007/s11207-011-9900-9.

Gordovskyy, M., Kontar, E.P. and Browning, P.K. (2016) Plasma motions and non-thermal line broadening in flaring twisted coronal loops, Astron. Astrophys. 589, A104, doi: 10.1051/0004-6361/201527249.

Gordovskyy, M., Browning, P.K. and Kontar, E. (2017) Polarisation of microwave emission from reconnecting twisted coronal loops. Astron. Astrophys. 604, A116. 10.1051/0004-6361/201629334

Gordovskyy,M., Browning, P.K. and Pinto,R.F. (2019) Combining MHD and kinetic modelling of solar flares, Adv. Space Res. 63, 1453-1465, doi:10.1016/j.asr.2018.09.024.

Gryaznevich,M.W. and Sykes, A. (2017) Merging-compression formation of high temperature spherical tokamak plasma, Nuc. Fus.  57, 072003, doi: 10.1088/1741-4326/aa4ffd.

Heyvaerts, J. and Priest, E.R. (1984) Coronal heating by reconnection in DC current systems - A theory based on Taylor's hypothesis  Astron. Astrophys 137, 63.

Hood, A.W. and Priest, E.R. (1979) Kink Instability of Solar Coronal Loops as the Cause of Solar Flares. Sol. Phys. 64, 303-321. doi:10.1007/BF00151441.

Hood, A.W., Browning, P.K. and van der Linden, R.A.M. (2009)
Coronal heating by magnetic reconnection in loops with zero net current,
Astron. Astrophys., 506,913-925,   doi = 10.1051/0004-6361/200912285

Hood, A.W., Browning, P.K., Cargill, P.J. and Tam, K.V. (2015) An MHD Avalanche in a Multi-threaded Coronal Loop, Astrophys. J. 817, 5, doi: 10.3847/0004-637X/817/1/5

Hussain, A.S., Browning, P.K. and Hood, A.W. (2017) A relaxation model of coronal heating in multiple interacting flux ropes. Astron. Astrophys. 600, A5. 10.1051/0004-6361/201629589

Hussain, A.S. (2018) Synergistic effects of neutrons and plasma on materials in fusion reactors and  relaxation of merging magnetic flux ropes in fusion and solar plasmas. Ph.D thesis (University of Manchester)

Ji, H. Prager, S. and Sarff, J. (1995) Conservation of Magnetic Helicity during Plasma Relaxation. Phys. Rev. Lett. 74, 2945-2948. 10.1103/PhysRevLett.74.2945

Kliem,B., T{\"o}r{\"o}k,T., Titov,V.S, Lionello,R., Linker,J.A, Liu,R., Liu,C.  and Wang,H (2014) Slow Rise and Partial Eruption of a Double-decker Filament. II. A Double Flux Rope Model. Astrophys. J. 792, 107, do:10.1088/0004-637X/792/2/107.

Kliem,B., Lee,J.,Liu,R.,White,S.M, Liu,C.  and Masuda,S (2021) Nonequilibrium Flux Rope Formation by Confined Flares Preceding a Solar Coronal Mass Ejection. Astrophys. J.909, 91, doi:10.3847/1538-4357/abda37.

Klimchuk, J. (2006) On solving the coronal heating problem. Sol. Phys. 234, 41-77, doi:10.1007/s11207-006-0055-z

Kontar, E.~P.,Hannah,I.~G. and Bian, N.~H (2011) Acceleration, Magnetic Fluctuations, and Cross-field Transport of Energetic Electrons in a Solar Flare Loop. Astrophys. J. Lett, 730, L22, doi:10.1088/2041-8205/730/2/L22

Kuridze, D., Mathioudakis, M., Kowalski, A.~F., Keys, P.~H., Jess, D.~B. Balasubramaniam, K.~S. and Keenan, F.~P.(2013). Failed filament eruption inside a coronal mass ejection in active region 11121. Astron. Astrophys., 552, A55.
doi: 10.1051/0004-6361/201220055

Leamon, R.J., Canfield, R.C., Blehm, Z.  and Pevtsov, A.A. (2003). 'What Is the Role of the Kink Instability in Solar Coronal Eruptions?', Astrophys.J.495, L255-L258

Liang, Y., Gimblett, C.G., Browning, P.K., Devoy, P., Koslowski, H.R., Jachmich, S. et al (2010) Multiresonance Effect in Type-I Edge-Localized Mode Control With Low n Fields on JET.  Phys. Rev. Lett.  105, 065001, doi:10.1103/PhysRevLett.105.065001.

Linton, M.G., Dahlburg,R.B. and Antiochos,S.K. (2001) Reconnection of Twisted Flux Tubes as a Function of Contact Angle. Astrophys. J. 553, 905-921, do:10.1086/320974.
 
Lionello, R, Velli, M., Einaudi, G. and Mikic, Z. (1998) Nonlinear Magnetohydrodynamic Evolution of Line-tied Coronal Loops. Astrophys. J. 494, 840-850. doi:10.1086/305221

Liu, R. (2020) Magnetic flux ropes in the solar corona; structure and evolution towards eruption. Res. Astron. Astrophys. 20, 165. doi:10.1088/1674-4527/20/10/165.

De Moortel, I. and Browning, P.K. (2015) Recent advances in coronal heating. Phil Trans Roy Soc A 373, 20140269-20140269, doi:10.1098/rsta.2014.0269.

McClements,K.G. (2019) Reconnection and fast particle production in tokamak and solar plasmas. Adv. Space  Res. 63,doi:10.1016/j.asr.2018.08.045~G.

Murray,S.A, Bloomfield,D.S. and Gallagher,P.T.(2013) Evidence for partial Taylor relaxation from changes in magnetic geometry and energy during a solar flare, Astron. Astrophys. 550, A119, doi:10.1051/0004-6361/201219964
        
Nandy,D., Hahn,M., Canfield,R.C. and Longcope,D.W. (2003)
Detection of a Taylor-like Plasma Relaxation Process in the Sun, Astrophys. J. Lett. 597, L73-L76, doi:10.1086/379815

Parker, E. N. (1988) Nanoflares and the Solar X-Ray Corona. Astrophys. J. 330, 474-479, doi:10.1086/166485.

Parnell,C.E. and De Moortel, I. (2012) A contemporary review of coronal heating.  Phil Trans Roy Soc A

Patsourakos,S, Vourlidas,A. and Stenborg,G. (2013) Direct Evidence for a Fast Coronal Mass Ejection Driven by the Prior Formation and Subsequent Destabilization of a Magnetic Flux Rope. Astrophys. J. 764, 125, doi:10.1088/0004-637X/764/2/125.

Patsourakos, S., Vourlidas, A., T{\"o}r{\"o}k,T.,Kliem, B. et al (2020) Decoding the Pre-Eruptive Magnetic Field Configurations of Coronal Mass Ejections. Space Sci. Rev. 216, 131, doi:10.1007/s11214-020-00757-9

Pinto,R.F., Vilmer,.N. and Brun, A.S.  (2015) Soft X-ray emission in kink-unstable coronal loops,  Astron. Astrophys 576, A37, doi:10.1051/0004-6361/201323358.

Pinto, R.F., Gordovskyy, M., Browning, P.K. and Vilmer, N. (2016) Thermal and non-thermal emission from reconnecting twisted coronal loops. Astron. Astrophys. 585, A159, doi:10.1051/0004-6361/201526633.

Priest, E.R. (2014) Magnetohydrodynamics of the Sun, Cambridge University Press doi:018A\&A...615A..84R.

Reid, J., Hood, A.W., Parnell, C.E., Browning, P.K. and Cargill, P.J. (2018) Coronal energy release by MHD avalanches: continuous driving. Astron. Astrophys. 615, A84, doi:10.1051/0004-6361/201732399.

Ripperda, B., Porth, O., Xia, C. and Keppens, R. (2017a) Reconnection and particle acceleration in interacting flux ropes - I. Magnetohydrodynamics and test particles in 2.5D, Mon. Not. Roy. Astro. Soc. 467, 3279-3298, doi:10.1093/mnras/stx379.

Ripperda, B., Porth, O., Xia, C. and Keppens, R. (2017b) Reconnection and particle acceleration in interacting flux ropes – II 3D effects on test particles in magnetically dominated plasmas, Mon. Not. Roy. Astro. Soc. 467, 3465-3482, doi:10.1093/mnras/stx1875.

Sharykin,I.N., Kuznetsov,A.A. and Myshyakov,I.I. (2018) Probing Twisted Magnetic Field Using Microwave Observations in an M Class Solar Flare on 11 February, 2014. Sol. Phys. 293, 34, doi:10.1007/s11207-017-1237-6.
 
Snow, B., Botha,G.J.J., Regnier, S., Morton, R.J., Verwichte, E. and Young, P.R. (2017) Observational Signatures of a Kink-unstable Coronal Flux Rope Using Hinode/EIS. Astrophys. J. 842, 16, doi:10.3847/1538-4357/aa6d0e.

Snow, B., Botha,G.J.J , Scullion, E., McLaughlin, J.A., Young, P.R. and Jaeggli, S.A. (2018) Predictions of DKIST/DL-NIRSP Observations for an Off-limb Kink-unstable Coronal Loop,
 Astrophys. J 863, 172, doi: 10.3847/1538-4357/aad3bc

Srivastava,A.K., Zaqarashvili,T.V, Kumar,P.  and Khodachenko,M.L. (2010) Observation of Kink Instability During Small B5.0 Solar Flare on 2007 June 4. Astrophys. J. 795, 292. doi:10.1088/0004-637X/715/1/292.
 
Stanier,A., Browning,P.K., McClements,K, Gordovskyy,M., Gryaznevich,M. and Lukin, V. (2013) Two fluid simulations of driven reconnection in the mega ampere spherical tokamak, Phys. Plas. 20, 122302, doi:10.1063/1.4830104

Tam, K.V., Hood, A.W., Browning, P.K. and Cargill,P.J. (2015)  Coronal heating in multiple magnetic threads. Astron. Astrophys 580, A122, doi:10.1051/0004-6361/201525995.

Tanabe.H.  Yamada,T. Watanabe, T.  et al (2017) Recent progress on magnetic reconnection research in the MAST spherical tokamak. Phys. Plas. 24, 056108, doi: 10.1063/1.4977922

Taylor, J.B. (1974) Relaxation of toroidal plasma and generation of reverse magnetic field. Phys Rev Lett. 30, 1139-1141,  doi:10.1103/PhysRevLett.33.1139

Taylor, J.B. (1984) Relaxation and magnetic reconnection in plasmas. Rev. Mod.Phys.58, 741-763, doi:10.1103/RevModPhys.58.741 

Taylor, J. B. (2000) Relaxation revisited. Phys. Plas. 7, 1623.  https://doi.org/10.1063/1.873984

Threlfall, J., Hood, A.W. and Browning, P.K. (2018) Flare particle acceleration in the interaction of twisted coronal flux ropes. Astron. Astrophys 611, A40, doi:10.1051/0004-6361/201731915

T{\"o}r{\"o}k,T., Kliem,B. and Titov,V.S. (2004) Ideal kink instability of a magnetic loop equilibrium, Astron. Astrophys. 413, L27-L30, doi: 10.1051/0004-6361:20031691

Williams,D.R, T{\"o}r{\"o}k,T., D{\'e}moulin,P., van Driel-Gesztelyi,L. and Kliem,B. (2005) Eruption of a Kink-unstable Filament in NOAA Active Region 10696, Astrophys. J. Lett. 628,L163-L166, doi:10.1086/432910

Woltjer, L. (1958) A Theorem on Force-Free Magnetic Fields. Proc. Nat. Acad. Sci. 44, 489-491, doi:10.1073/pnas.44.6.489

Yamada,M,, Ono,Y., Hayakama,A., Katsurai,M. and  Perkins,F.W. (1990) Magnetic reconnection of plasma toroids with cohelicity and counterhelicity. Phys. Rev. Lett. 65, 721.

Yang,K., Guo,Y. and Ding,M.D. (2016) Quantifying the Topology and Evolution of a Magnetic Flux Rope Associated with Multi-flare Activities. Astrophys. J. 824, 148. doi:10.3847/0004-637X/824/2/148

Yeates,A.M., Russell, A.J.B. and Hornig,G (2015) Physical role of magnetic constraints in localized magnetic relaxation. Proc. Roy. Soc. A 471, 20150012, doi:10.1098/rspa.2015.0012 

\end{document}